\documentclass[reprint,superscriptaddress,showpacs,showkeys,preprintnumbers,amsmath,amssymb,aps,prb]{revtex4-1}
\usepackage{graphicx}    
\usepackage{float}
\usepackage{color}

\begin{document}

\title{Magnetic properties and Curie temperatures of disordered Heusler compounds: Co$_{1+x}$Fe$_{2-x}$Si}

\author{Julia~Erika~Fischer}
\affiliation{Max Planck Institute for Chemical Physics of Solids, D-01187 Dresden, Germany}

\author{Julie~Karel}
\email{julie.karel@cpfs.mpg.de}
\affiliation{Max Planck Institute for Chemical Physics of Solids, D-01187 Dresden, Germany}

\author{Simone~Fabbrici}
\affiliation{Institute of Materials for Electronics and Magnetism, I-43124 Parma, Italy}

\author{Peter Adler}
\affiliation{Max Planck Institute for Chemical Physics of Solids, D-01187 Dresden, Germany}

\author{Siham~Ouardi}
\affiliation{Max Planck Institute for Chemical Physics of Solids, D-01187 Dresden, Germany}

\author{Gerhard~H.~Fecher}
\affiliation{Max Planck Institute for Chemical Physics of Solids, D-01187 Dresden, Germany}

\author{Franca~Albertini}
\affiliation{Institute of Materials for Electronics and Magnetism, I-43124 Parma, Italy}

\author{Claudia~Felser}
\affiliation{Max Planck Institute for Chemical Physics of Solids, D-01187 Dresden, Germany}

\date{\today}

%%%%%%%%%%%%%%%%%%%%%%%%%%%%%%%%%%%%%%%%%%%%%%%%%%%%%%%%%%%%%%%%%%%%%%%%%%%%%%%
\begin{abstract}

The local atomic environments and magnetic properties were investigated for a
series of Co$_{1+x}$Fe$_{2-x}$Si (0~$\leq$~$x$~$\leq$~1) Heusler compounds. While the total magnetic moment in these compounds increases
with the number of valance electrons, the
highest Curie temperature ($T_C$) in this series
was found for Co$_{1.5}$Fe$_{1.5}$Si, with a $T_C$ of 1069~K (24~K higher than the well known Co$_2$FeSi).
$^{57}$Fe M{\"o}ssbauer spectroscopy was used to characterize the local atomic order and to estimate the
Co and Fe magnetic moments. Consideration of the local magnetic moments and the exchange integrals is necessary to understand the trend in $T_C$.

\end{abstract}
%%%%%%%%%%%%%%%%%%%%%%%%%%%%%%%%%%%%%%%%%%%%%%%%%%%%%%%%%%%%%%%%%%%%%%%%%%%%%%%

\pacs{75.30.Cr, 76.80.+y, 75.50.Bb, 85.75.-d}

\maketitle

%%%%%%%%%%%%%%%%%%%%%%%%%%%%%%%%%%%%%%%%%%%%%%%%%%%%%%%%%%%%%%%%%%%%%%%%%%%%%%%
\section{Introduction}

Co$_2$-based Heusler compounds belong to one of the most promising classes of
materials for high performance spintronic
devices because they have been predicted to be
half-metallic ferromagnets (HMF)~\cite{CoHMF1,KAW16,KAS14,HAZ16}. These materials exhibit complete spin
polarization, arising from a band gap at the Fermi energy for the minority
electrons and metallic-like behavior of majority
carriers~\cite{HMF1,HMF2,HMF3,HMF4}. This high spin polarization has been
recently demonstrated experimentally~\cite{SPexp}. The compounds with high magnetic moments also exhibit high Curie temperatures,
for example, well ordered Co$_2$FeSi has a magnetic moment of $6\:\mu_B$ and a
$T_C$ of 1040~K~[\onlinecite{am2}], making these materials important for applications.  High Curie temperatures in HMF are desirable for stable spintronic devices
because they reduce magnetic fluctuations at the working temperatures that are
usually above room temperature. 

\begin{figure}[htb]
    \centering
    \includegraphics[width=7cm]{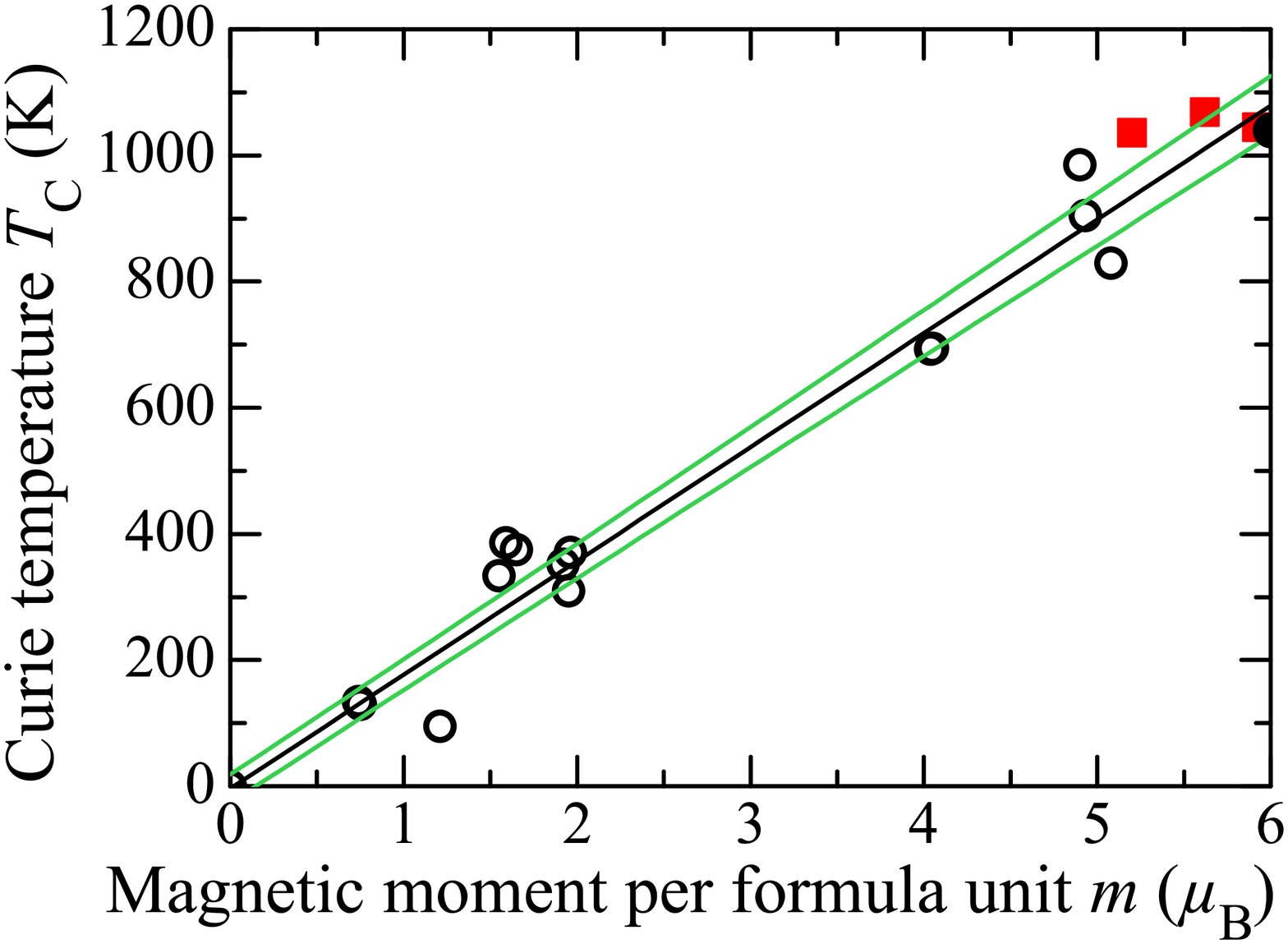}
    \caption{(Color online) Dependency of the $T_C$ on the magnetic moment for Co$_2$-based Heusler compounds.\\
              Data for open circles are collected from Reference~[\onlinecite{Syn1}]. The closed circle data 		      	     point at $m~=~6\:\mu_{\rm B}$ is taken from Reference~[\onlinecite{am2}]. Closed
              squares are from the measurements reported in this work.
              The lines result from a fit of the cited data to a linear model and give the expectation in the 95\% confidence interval. 
              }
\label{fig:old}
\end{figure}

Previous studies have shown that the total magnetic moment of Co$_2$-based 
half-metallic ferromagnetic Heusler compounds increases linearly over a large range
of the number of valence electrons ($N_{\rm{VE}}$). This linear dependence is
derived from the well-known Slater--Pauling rule~\cite{HMF4, Kubler, CoHMF2}.
Further, it was found that the Curie temperatures of these materials increase with
increasing magnetic moment~\cite{m-Tc,CoHMF2}, with a dependence that seems to be
nearly linear over a wide range of magnetic moments (see Figure~\ref{fig:old}).
In a simple spin--only molecular field approach, $T_C$ should scale with the
square of the spin ($T_C\propto S(S+1)$). It was thus suggested that the
linearity in $m$, which is proportional to $S$, is caused by modifications in the Heisenberg
exchange coupling coefficients~\cite{CoHMF2}. Figure~\ref{fig:old} shows a plot
of known Curie temperatures of Co$_2$ based Heusler compounds. It is evident
that most of the data points are outside of the 95\% confidence bounds. The
statistical significance for a linear model may thus not be very high.
In fact, deviations from the linear trend have already been discussed in 
Reference~[\onlinecite{m-Tc}], where it was shown theoretically that in particular compounds with 27 valence electrons
exhibit significantly lower Curie temperatures resulting from a drastic change in the
exchange coupling coefficients at this valence electron concentration.
Based on the electronic structure calculations for compounds with more than 27 valence electrons, it was found that the exchange
average only slightly increases, and the upward trend in the Curie temperatures is
due to the large local moments. In other cases,
the measured and calculated Curie temperatures and magnetic moments deviate from
each other because of disorder in the investigated samples that is not accounted for
in the theoretical approaches. It is further seen from Figure~\ref{fig:old} that
large deviations from an increasing trend are found in small regions around a
selected magnetic moment where one actually may observe a decrease. Therefore,
it is worth analyzing the behavior of the Curie temperatures in a small range
of magnetic moments and investigating the influence of disorder on the $T_C-m$
relation.  The present work investigates Heusler compounds
that are known to exhibit extremely high Curie temperatures~\cite{Fe2CoSi,am2},
namely Co$_{1+x}$Fe$_{2-x}$Si. The valence electron concentration varies by only
one electron in this series where $x$ is between 0 and 1; therefore, the magnetic moments
vary also by just $1\:\mu_B$.

%%%%%%%%%%%%%%%%%%%%%%%%%%%%%%%%%%%%%%%%%%%%%%%%%%%%%%%%%%%%%%%%%%%%%%%%%%%%%%%
\section{Experimental Methods}

Polycrystalline Co$_{1+x}$Fe$_{2-x}$Si samples with $x$~=~0, 0.2,
0.4, 0.5, 0.6, and 1.0 were synthesized by arc melting. Stoichiometric amounts of the elements
(purity of 99.999\% or higher) were arc melted on a water cooled copper hearth
under argon atmosphere. The ingots were remelted four times to improve the
samples' homogeneity. They were put in a welded tantalum tube, and annealing was
carried out in a sealed quartz ampoule, followed by rapid quenching in a mixture
of ice and water. Co$_2$FeSi was annealed at 1300~K for 21~days, CoFe$_2$Si was
annealed at 1073~K for 4~days, and the samples with intermediate stoichiometry were annealed
at 1000~K for 7~days to the homogeneity and crystalline structures~\cite{Syn1,Jun08,Syn3}.

The composition and phase homogeneity of the samples were verified by
wavelength dispersive X-ray spectroscopy (WDS) using a Cameca SX 100 electron
microprobe.
Wavelength dispersive X-ray spectroscopy revealed that obtained
and nominal target compositions are in good agreement within the experimental uncertainty
(see Table~\ref{tab:comp}). Therefore, the compounds will be described in the following by their
nominal compositions.

The structure was determined and the phase homogeneity further verified by
powder X-ray diffraction using an image plate Huber G670 Guinier camera with a
Ge(111) monochromator and Co~K$_\alpha$ radiation. To collect the data, the
samples were evenly distributed on Mylar foil and a pattern in the range of
$2\theta=10^{\circ} -100^{\circ}$ was recorded at room temperature. For
determination of the lattice parameters, the silicon standard reference
material{\textregistered} 640c was added. Indexing and refinement of the lattice
parameter was done with the programs WIN XPOW~\cite{Win} and
PowderCell~\cite{PC}.

Magnetostructural investigations were performed at room temperature using
$^{57}$Fe M{\"o}ssbauer spectroscopy on a standard WissEl spectrometer operated
in the constant acceleration mode and equipped with a $^{57}$Co/Rh source. The
powdered samples containing approximately 10~mg/cm$^{2}$ of Fe were diluted with
boron nitride to ensure a homogeneous distribution. Acrylic glass sample
containers were used. All isomer shifts are given relative to $\alpha$-iron. The
data were evaluated with the MossWinn\cite{mbfit} program using the thin
absorber approximation. Hyperfine field distributions were derived using the
Hesse--R{\"u}bartsch method implemented in MossWinn.

The magnetization and hysteresis were measured with a SQUID magnetometer
(Quantum Design; MPMS~3 or MPMS~XL7). The measurements were carried out at 1.8~K
in induction fields in the range of -5 to +5~T. Temperature dependent AC
susceptibility measurements were performed using thermomagnetic analysis (TMA)
in order to determine the Curie temperature of the samples. Data were collected
in a home-built susceptometer with a working temperature range from room
temperature to 1373~K. An alternating magnetic induction field of 0.4~mT was
applied. The instrumental precision is $\pm2$~K derived from the sensor's
accuracy and calibration with an iron standard sample. A description of the TMA
principle is found in Reference~[\onlinecite{TMA}].

%%%%%%%%%%%%%%%%%%%%%%%%%%%%%%%%%%%%%%%%%%%%%%%%%%%%%%%%%%%%%%%%%%%%%%%%%%%%%%%
\section{Results}

%%%%%%%%%%%%%%%%%%%%%%%%%%%%%%%%%%%%%%%%%%%%%%%%%%%%%%%%%%%%%%%%%%%%%%%%%%%%%%%
\subsection{Structural Analysis}

\begin{figure}[htb]
    \centering
    \includegraphics[width=7cm]{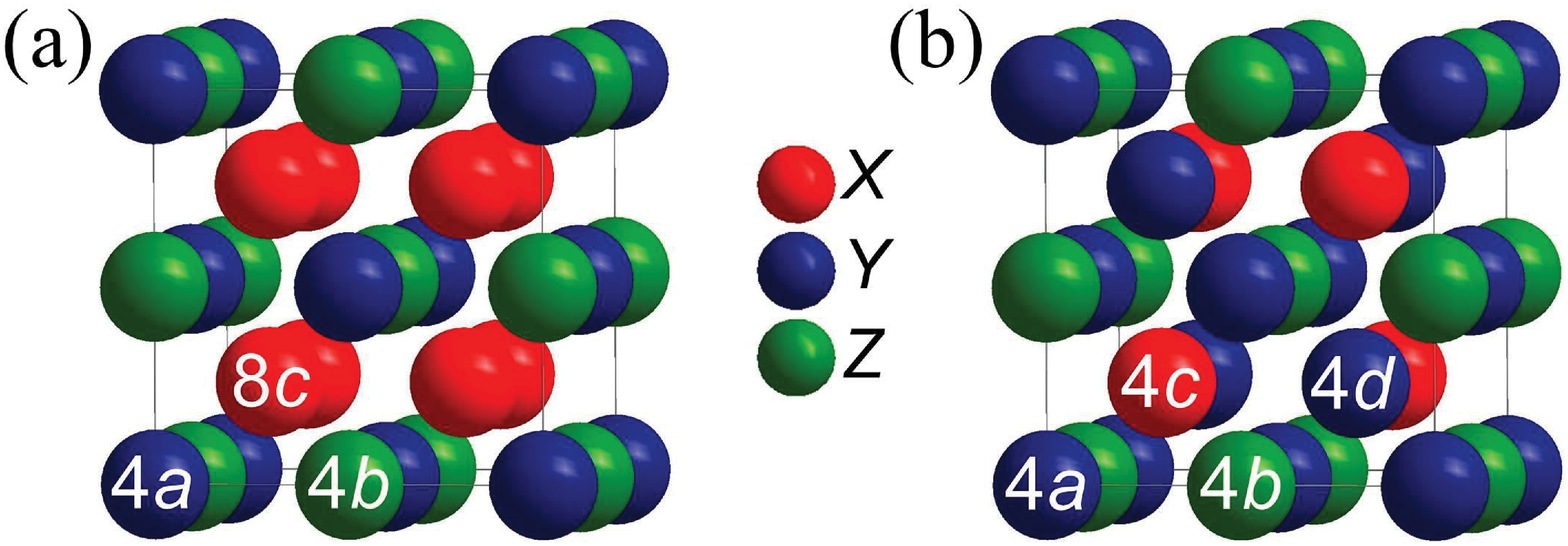}
    \caption{(Color online) Crystal structures of Co$_2$FeSi and Fe$_2$CoSi.\\
             Shown are:
             (a) the regular Heusler $L2_{1}$ structure (space group 225) and
             (b) the inverse $X$ structure (space group 216).
             The atoms are labeled by the corresponding Wyckoff positions. }
\label{fig:struc1}
\end{figure}

Figure~\ref{fig:struc1} shows the two different structure types that occur for
ordered Heusler compounds generally. The two structure types are the regular
$L2_1$ type (Cu$_2$MnAl, cF16, $F\:m\overline{3}m$, 225) and the inverse $X$
type (Li$_2$AgSb, cF16, $F\:\overline{4}3m$, 216). In the Co$_{1+x}$Fe$_{2-x}$Si
series, the general atoms assigned in Figure~\ref{fig:struc1} by $X$, $Y$, and
$Z$ correspond to Co, Fe, and Si, respectively.

\begin{figure}[htb]
    \centering
    \includegraphics[width=7.5cm]{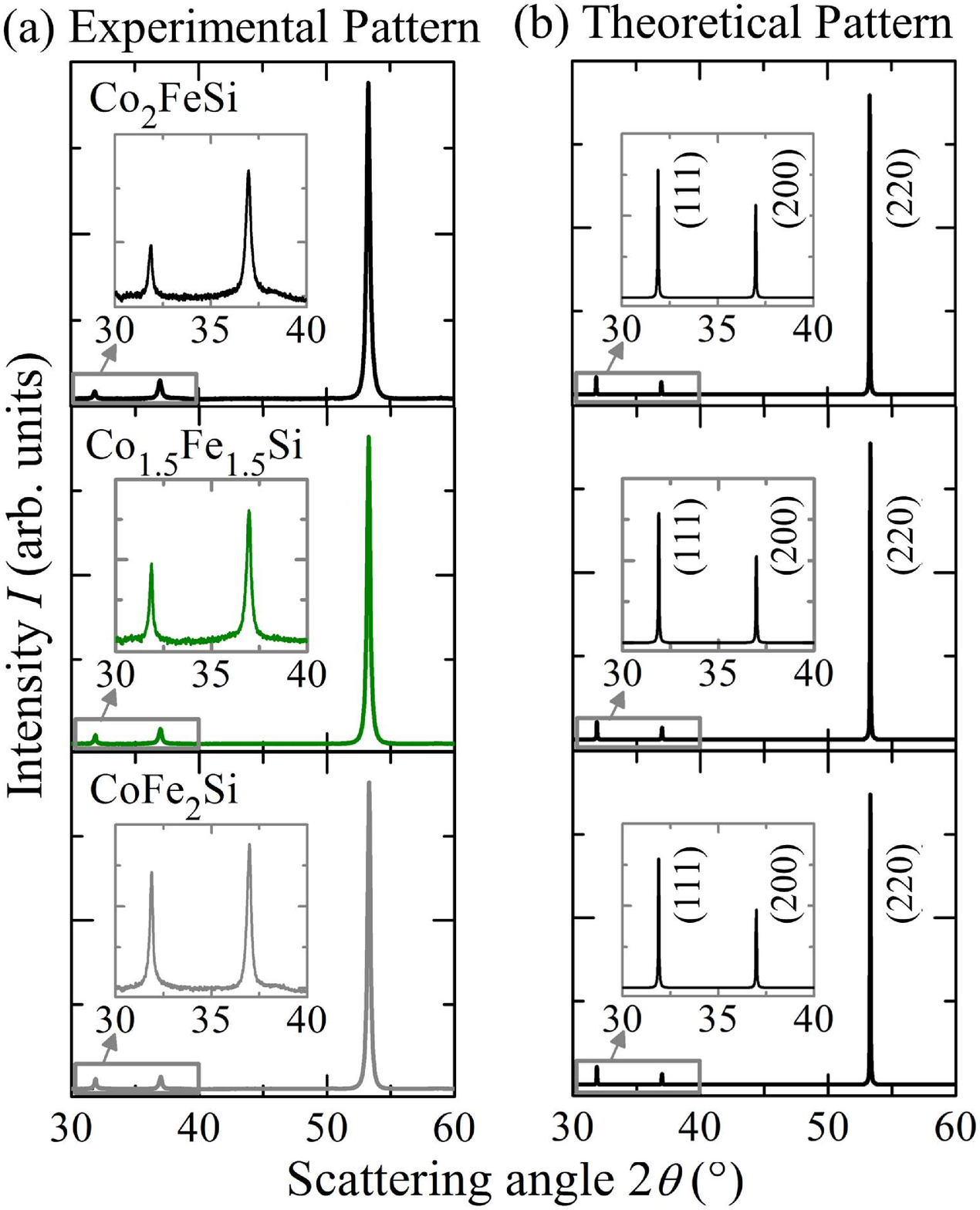}
    \caption{(Color online) Powder XRD of selected Co$_{2-x}$Fe$_{1+x}$Si samples \\
             Representative (a) experimental and (b) theoretical X-ray diffraction patterns
             of Co$_2$FeSi, Co$_{1.5}$Fe$_{1.5}$Si and CoFe$_2$Si.
             The theoretical patterns were calculated for the well ordered structures with Co replacing
             Fe on the 4d position of the $X$ type structure. 
             }
\label{fig:XRD}
\end{figure}

Selected results from powder XRD are shown in Figure~\ref{fig:XRD}. The
diffraction patterns were indexed using a face centered cubic cell
(Figure~\ref{fig:XRD}). The lattice constant is $a=5.64$~{\AA} for all samples.
This corresponds to the lattice parameters found in other
works~\cite{CFSTc2,Nicu,am2,Fe2CoSi}. Major changes of the lattice parameters
are not expected within the composition of the series because the radii of Co
and Fe differ by about 0.01~{\AA} only~\cite{Green}.

In the X-ray diffraction pattern of Co$_2$FeSi traces of a minority phase (whose
estimated weight fraction is less than 1\%) are deduced by the presence of
extremely small peaks around $2\theta=51^\circ$ and $2\theta=9^\circ$. The
investigations discussed below in Sections~\ref{sec:magprop}
and~\ref{sec:moessbauer} will show that these minor impurities do not affect the
magnetic properties of the sample. The X-ray diffraction patterns of the other
samples do not indicate secondary phases.

An analysis of the intensities of the diffraction peaks shows that the theoretical
patterns of the corresponding ordered structures (Figure~\ref{fig:XRD}, right)
exhibit different intensity ratios than the experimental patterns. Corresponding
values are given in Table~\ref{tab:comp}. This result demonstrates that disorder is present in
the samples.
A detailed Rietveldt analysis reveals that the Co$_2$FeSi sample exhibits a
large degree of Fe-Si anti side disorder and is composed of 80\% $L2_1$
and 20\% $B2$ type phases. The type of disorder in Fe$_2$CoSi is not easily
distinguished; it is likely composed of about 30 to 35\% of a $B2$ type phase
with the remaining phase adopting either an  $L2_1$ type or $X$ type structure.
A determination of the type of disorder is even further complicated in the mixed
compounds due to the relatively similar scattering parameters of Fe and Co. 
More details of the disordered structures and the local environments of
the Fe atoms will be discussed in Section~\ref{sec:moessbauer}, which presents results
from $^{57}$Fe M{\"o}ssbauer spectroscopy.

\begin{table*}[htb]
  \caption{\label{tab:comp}Results of the phase analysis of the alloys
           Co$_{1+x}$Fe$_{2-x}$Si and representative intensity ratios derived from the
           experimental and theoretical X-ray diffraction patterns for Co$_2$FeSi,
           CoFe$_2$Si and Co$_{1.5}$Fe$_{1.5}$Si from the patterns shown in Figure~\ref{fig:XRD}.
           The relative error of the composition is 0.30\% for Si and
           0.45\% for Co and Fe for the values derived from WDS, which leads to the same
           uncertainty in the number of valence electrons.}
  \centering
    \begin{ruledtabular}
      \begin{tabular}{llllllllcccccc}
      $x$ & $N_v$ & \multicolumn{3}{l}{target composition} & \multicolumn{3}{l}{WDS composition} & \multicolumn{3}{c}{experimental XRD} & \multicolumn{3}{c}{theoretical XRD}\\
      &  & Co & Fe & Si & Co & Fe & Si & $I_{111}/I_{200}$ & $I_{111}/I_{220}$ & $I_{200}/I_{220}$ & $I_{111}/I_{200}$ & $I_{111}/I_{220}$ & $I_{200}/I_{220}$\\
      \hline
      0   & 29     & 1.00 & 2.00 & 1.00   & 1.01 & 2.00 & 0.99   & 0.53 & 0.02 & 0.03   & 1.64 & 0.06 & 0.04 \\
      0.2 & 29.2   & 1.20 & 1.80 & 1.00   & 1.20 & 1.81 & 0.99   &-     &-     &-       &-     &-     &-     \\
      0.4 & 29.4   & 1.40 & 1.60 & 1.00   & 1.40 & 1.61 & 0.99   &-     &-     &-       &-     &-     &-     \\
      0.5 & 29.5   & 1.50 & 1.50 & 1.00   & 1.51 & 1.51 & 0.98   & 0.50 & 0.02 & 0.04   & 1.49 & 0.06 & 0.04 \\
      0.6 & 29.6   & 1.60 & 1.40 & 1.00   & 1.60 & 1.41 & 0.99   &-     &-     &-       &-     &-     &-     \\
      1   & 30     & 2.00 & 1.00 & 1.00   & 1.99 & 1.01 & 1.00   & 0.25 & 0.01 & 0.05   & 1.37 & 0.06 & 0.04 \\
      \end{tabular}
    \end{ruledtabular}
\end{table*}

%%%%%%%%%%%%%%%%%%%%%%%%%%%%%%%%%%%%%%%%%%%%%%%%%%%%%%%%%%%%%%%%%%%%%%%%%%%%%%%
\subsection{Magnetic Properties}
\label{sec:magprop}

The field dependency of the magnetization was studied at 1.8~K, and the behavior is
typical of soft magnetic materials. The saturation magnetization per formula
unit has values between 5 and $6\:\mu_{\rm B}$. The moment of Co$_2$FeSi is
$5.92\:\mu_{\rm{B}}$, which is in good agreement with previous work
($5.91\:\mu_{\rm B}$ at 10~K~[\onlinecite{Nicu}] and $5.97\:\mu_{\rm B}$ at
5~K~[\onlinecite{am2}]). The Slater--Pauling rule describes the dependence of
the magnetic moment on the valence electron concentration. For ordered,
half-metallic ferromagnetic Heusler compounds this dependence is given
by~\cite{CoHMF2}:

\begin{equation}
  m = N_v - 24.
\end{equation}

The experimentally determined values follow this equation only roughly
(Figure~\ref{fig:Tc}~(a)). The slightly higher values found for $m$ in the Fe rich
members of this series are not unusual. Similar deviations from the theoretical
values have been reported previously for Heusler compounds~\cite{CoHMF2}.

\begin{figure}[htb]
\centering
\includegraphics[width=7cm]{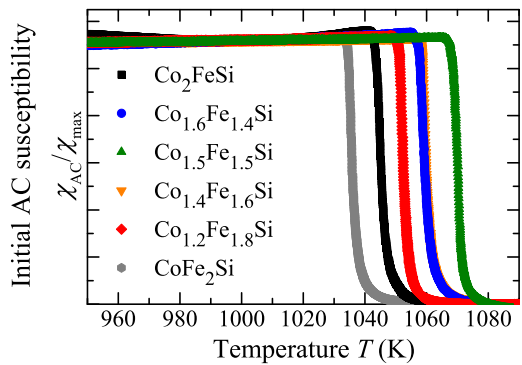}
   \caption{(Color online) Heating curves of the AC susceptibility measurements near the Curie point.}
\label{fig:TMA}
\end{figure}

The Curie temperatures were determined by measuring the AC susceptibility. The
results are shown in Figure~\ref{fig:TMA}. In this experiment, the temperature
dependence of the initial AC susceptibility ${\chi}_{\rm{AC}}$ provides the
magnetic critical temperatures of the samples. Since the recorded AC
susceptibility is influenced by many extrinsic features, such as microstructure,
grain texture and lattice defects, the measurements shown in
Figure~\ref{fig:TMA} were normalized to the maximum values. The shallow dip that
is observed in some measurements below the Curie temperature is due to the fact
that the samples were not measured at saturation but in a field of only 0.4~mT.

The Curie temperatures are determined from the inflection point of the
registered curves. The width of the transition depends on the applied
magnetic field and sample homogeneity. The full width at half maximum of the
first derivative of the curve as well as the instrumental precision were taken
into account for determining the accuracy of the obtained Curie temperatures.
Hence, the $T_C$ of Co$_2$FeSi is 1045~K with an uncertainty of 0.4\%.
This value compares well with the inflection point of the magnetization versus
temperature measurements reported in Reference~[\onlinecite{CFSTc1}].

The $T_C$ of the series changes with iron content, where it reaches a maximum
for $x=1/2$ (see Figure~\ref{fig:Tc}~(b)). The maximum Curie
temperature of 1069~K found for Co$_{1.5}$Fe$_{1.5}$Si is 24~K higher than the
value derived for Co$_2$FeSi.

%%%%%%%%%%%%%%%%%%%%%%%%%%%%%%%%%%%%%%%%%%%%%%%%%%%%%%%%%%%%%%%%%%%%%%%%%%%%%%%
\subsection{$^{57}$Fe M{\"o}ssbauer Spectroscopy}
\label{sec:moessbauer}

\begin{figure}[htb]
\centering
   \includegraphics[width=7cm]{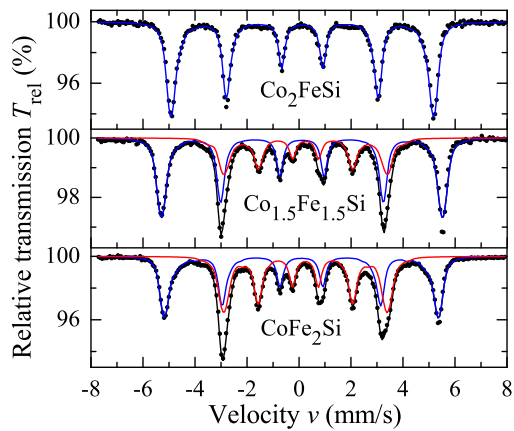}
   \caption{(Color online) Representative $^{57}$Fe M{\"o}ssbauer spectra of Co$_{1+x}$Fe$_{2-x}$Si collected at room temperature.
            Solid lines correspond to the calculated spectra obtained from fits assuming distributions of hyperfine fields.
            The red and blue lines correspond to the subspectra for the two different iron sites.}
\label{fig:mb}
\end{figure}

$^{57}$Fe M{\"o}ssbauer spectroscopy is well suited to unravel the atomic order
in Co$_{1+x}$Fe$_{2-x}$Si Heusler alloys~\cite{Jun08}. The M{\"o}ssbauer spectra
of the compounds are composed of hyperfine sextets as expected for magnetically
ordered phases (representative spectra are shown in Figure~\ref{fig:mb}). All
spectra exhibit some line broadening in addition to the linewidth of the
$\alpha$-Fe standard and also some asymmetry. Therefore, the spectra were evaluated with
distributions of hyperfine fields $B_{\rm{hf}}$ rather than with single
sextets of Lorentzian shape. The results of the data analysis are summarized
in Table \ref{tab:mb}, where $IS$ corresponds to the isomer shifts and
$<$$B_{\rm{hf}}$$>$ to the average hyperfine field values of the distribution.
$B_{\rm{hf}}$ is the hyperfine field with the highest probablity.

The spectrum of Co$_2$FeSi was fitted by a single hyperfine field distribution
(component~I). The $B_{\rm{hf}}$ of 31.1~T is in good agreement with Fe atoms
located on the $4a$ position of a regular Heusler structure. Conversely, the
spectrum of CoFe$_2$Si features a second six-line pattern (component~II) with a
considerably smaller $B_{\rm{hf}}$. The areas of the subspectra are nearly
equal, which verifies that CoFe$_2$Si adopts a Heusler structure where half of
the Co atoms at the $8c$ sites is replaced by Fe atoms.

\begin{figure}[htb]
\centering
   \includegraphics[width=7.5cm]{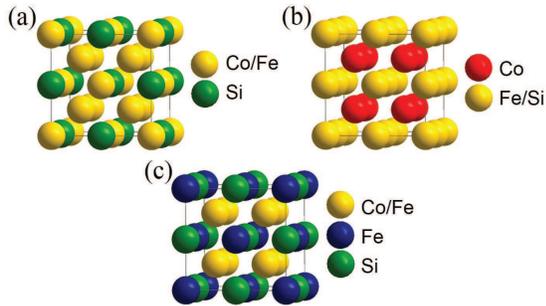}
   \caption{(Color online) Examples for the possible disorders occuring in the series Co$_{1+x}$Fe$_{2-x}$Si\\
             (a) D0$_{3}$-type disorder between Fe and Co on the $8c$ and $4a$ Wyckoff positions;
             (b) B2-type disorder between Fe and Si on the $4a$ and $4b$ Wyckoff positions;
             (c) L2$_{1b}$-type disorder between Co and Fe on the $8c$ Wyckoff positions. }
\label{fig:struc2}
\end{figure}

Due to the different local environments of the two iron sites, their local
magnetic moments $m_{\rm{Fe}}$ and thus their $B_{\rm{hf}}$ values are
different. This fact allows unambiguous identification of the two different iron
sites. Increasing the iron content in Co$_{1+x}$Fe$_{2-x}$Si leads to a
continuous increase of the relative intensity of component~II, verifying that
the additional Fe atoms successively replace the Co atoms on the $8c$ positions
of the regular structure. The relative area fractions of component~II correspond
reasonably well to the relative fractions of Fe atoms on the former $8c$ sites
as calculated from the nominal stoichiometry (Table~\ref{tab:mb},
Fe($8c$)$_{area}$ and Fe($8c$)$_{calc.}$, respectively). These results confirm
the high degree of atomic order with respect to occupation of the $4a$ and
former $8c$ sites by iron. In particular, there is no significant interchange
between Fe on the $4a$ sites and Co, which would lead to a decreased area
fraction of component~I in the M{\"o}ssbauer spectra. Therefore, disorder as
shown in Figure~\ref{fig:struc2}(a) can be excluded.

\begin{figure}[htb]
\centering
\includegraphics[width=6.8cm]{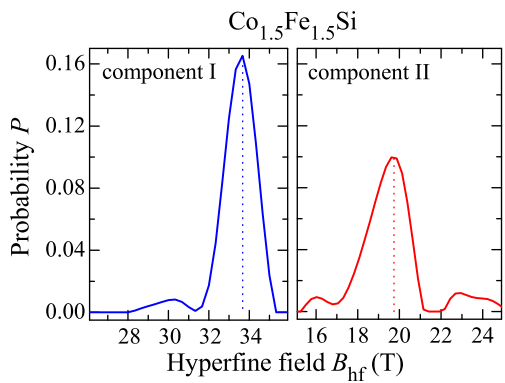}
\caption{(Color online) Hyperfine field distributions for components~I and II of Co$_{1.5}$Fe$_{1.5}$Si
          obtained from the data analysis of the corresponding M{\"o}ssbauer spectrum (Figure~\ref{fig:mb}).
          The labels I and II correspond to the Fe atoms on the $4a$ and former $8c$ sites, respectively.
          The dotted lines indicate the most probable $B_{\rm{hf}}$ values given in Table~\ref{tab:mb}.}
\label{fig:mb2}
\end{figure}

On the other hand, the $B_{\rm{hf}}$ distributions show a broadening of about
$2-3$~T and some additional asymmetry. As a representative example, the
$B_{\rm{hf}}$ distributions for $x=1/2$ are shown in Figure~\ref{fig:mb2}.
All data show a second weak feature for component~I with $B_{\rm{hf}}$ being
about 3~T smaller than that of the main band, which leads to a slight asymmetry
at the lower velocity value side of the peaks of the outer sextet (blue curve in
Figure~\ref{fig:mb}). This may indicate a small degree of Fe($4a$)~--~Si($4b$)
disorder (see Figure~\ref{fig:struc2}(b)), which does not change the nearest
neighbor environment of iron and thus modifies $B_{\rm{hf}}$ only slightly.
The $B_{\rm{hf}}$ distribution of component~II is even broader and reveals
some asymmetry as well, indicating some Fe~--~Co disorder on the former $8c$
positions. This corresponds to disorder of the kind shown in
Figure~\ref{fig:struc2}(c).

\begin{figure}[htb]
\centering
  \includegraphics[width=7.5cm]{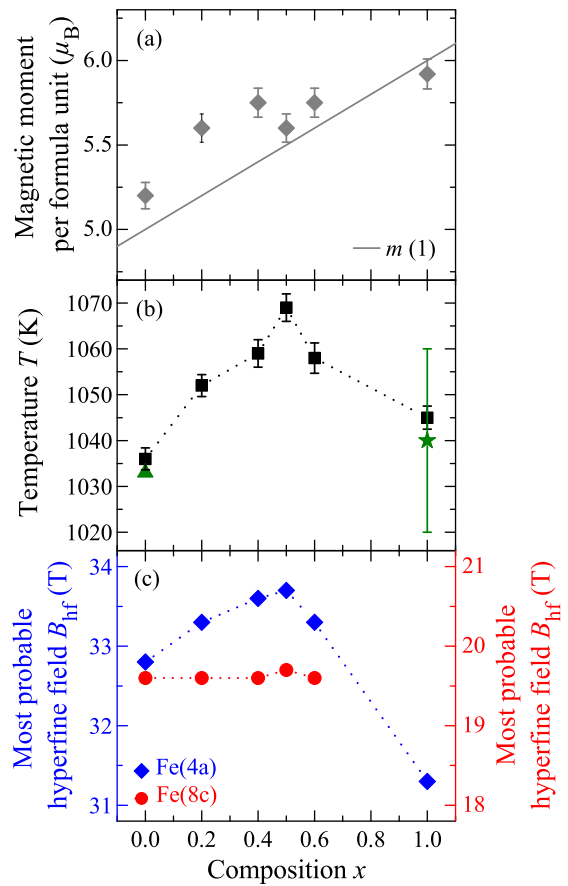}
  \caption{(Color online)
         (a) Trend of the magnetic moment per formula unit $m$ as a function of the composition $x$.
         The solid line corresponds to the moment calculated from equation~(1).
         (b) The Curie temperatures of the Co$_{1+x}$Fe$_{2-x}$Si-series.
         The star is the value for the inflection point from the magnetic measurement of Reference~[\onlinecite{am2}]
         and the triangle is the value extracted from the inflection point of the curves (with permission)
         measured by Du {\it et al.}~\cite{Fe2CoSi} (no error bars were given).
         (c) The most probable hyperfine fields of iron on the $4a$ and former $8c$ Wyckoff positions
         of the Co$_{1+x}$Fe$_{2-x}$Si-series as functions of $x$. Dotted lines are to guide the eye.}
\label{fig:Tc}
\end{figure}

The present spectra confirm the trends in the isomer shifts and hyperfine fields
reported previously~\cite{Jun08, kseno}. In particular, that both the most
probable $B_{\rm{hf}}$ value and the $IS$ of component~I depend slightly on the
composition and thus on $N_v$. Whereas $IS$ increases with $N_v$, a broad
maximum near $N_v=29.5$ (Table~\ref{tab:mb}) is observed for $B_{\rm{hf}}$.
Remarkably, the trend in $B_{\rm{hf}}$ of component~I is similar to the trend in
the Curie temperatures (Figure~\ref{fig:Tc}~(b), (c)). On the other hand, both, $IS$ and
$B_{\rm{hf}}$ of component~II are nearly constant. The small variation of the
hyperfine fields within the Co$_{1+x}$Fe$_{2-x}$Si series indicates that the
changes in the iron magnetic moments in response to the change in valence
electron concentration are small. It is rather the cobalt moment which is
altered.

\begin{figure}[htb]
\centering
   \includegraphics[width=7.5cm]{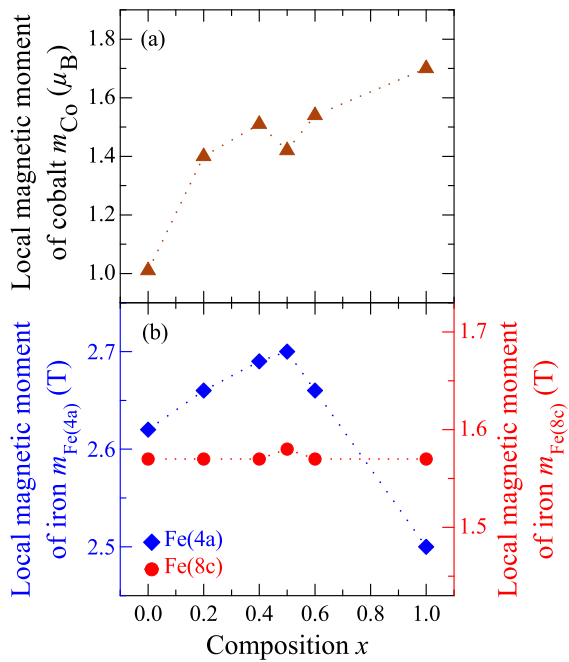}
   \caption{(Color online)
            (a) The local magnetic moment of the Co atoms and
            (b) the local magnetic moments of the the different Fe atoms as estimated from $B_{\rm{hf}}$ as functions of $x$.
            Dotted lines are to guide the eye.}
\label{fig:m}
\end{figure}

The M{\"o}ssbauer data may be used to estimate the local magnetic moments when
it is assumed that the magnetic hyperfine field is proportional to the ordered
magnetic moment ${m}_{\rm{Fe}}$  at the corresponding iron site. Choosing the
effective hyperfine coupling constant
$A~=~12.5$~T/$\mu_{\rm B}$~[\onlinecite{MAS79}] as in previous work on
Heusler phases Fe$_{2}YZ$~[\onlinecite{Gasi2013}] and assuming that $A$ does not
change within the limited range of compositions considered here, one obtains the
magnetic moments on the two iron sites given in Table~\ref{tab:mb}. The moments
on the $4a$ sites are about $1\:\mu_{\rm B}$ larger than those on the former
$8c$ sites.

Additionally, considering the total magnetic moment from the
magnetization studies the local magnetic moment ($m_{\rm{Co}}$) of the Co atoms
can be estimated. This analysis suggests that the increase in the total magnetic
moment with increasing valence electron concentration mainly results from an
increased local magnetic moment at the Co sites (Table~\ref{tab:mb}). The slight
variation of the hyperfine fields at the $4a$ sites then may be attributed to a
slight modulation of the magnetic moment by about $0.2\:\mu_{\rm B}$ within the
Co$_{1+x}$Fe$_{2-x}$Si series, whereas the magnetic moments of the iron atoms
entering the former $8c$ sites are essentially independent of $x$.

The present analysis yields reasonable estimates for the local moments of
Co$_{2}$FeSi and CoFe$_{2}$Si. The estimated moments of $2.5\:\mu_{\rm B}$ and
$1.7\:\mu_{\rm B}$ for Fe and Co are in good agreement with the corresponding
moments of $2.4\:\mu_{\rm B}$ and $1.7\:\mu_{\rm B}$ derived from a neutron
diffraction study of Co$_{2}$FeSi~[\onlinecite{NIC79}]. In the case of CoFe$_{2}$Si, a
moment of $2.6\:\mu_{\rm B}$ for the Fe $4a$ sites is obtained from both the
present M{\"o}ssbauer analysis and the neutron diffraction data~\cite{NIC79}.
However, for all alloys containing cobalt as well as iron atoms on the $8c$-like
sites, the corresponding magnetic moments obtained from neutron diffraction are
averages which cannot be decomposed without further assumptions. Niculescu {\it et
al.}~\cite{NIC79} \emph{assumed} that $m_{\rm{Co}}$ is \emph{constant}
($=1.7\:\mu_{\rm B}$) within the whole series of Fe$_{3-y}$Co$_{y}$Si with
$y \leq 2$ (which corresponds to $-1~\leq~x~\leq~+1$ in our notation
Co$_{1+x}$Fe$_{2-x}$Si). With this assumption the iron magnetic moments at the
$8c$-like sites were calculated and found to vary with composition. However,
electronic structure calculations on CoFe$_2$Si\cite{LUO11} as well as on the
germanium analog CoFe$_2$Ge~[\onlinecite{Gasi2013,REN10}] all yield local cobalt moments
of about $1\:\mu_{\rm B}$, in agreement with our analysis of the M{\"o}ssbauer
data of CoFe$_2$Si (Table~\ref{tab:mb}) and CoFe$_2$Ge~[\onlinecite{Gasi2013}]. The
electronic structure calculations support the observation that it is mainly the
cobalt moment that increases with increasing valence electron concentration.

In a work critically discussing the estimation of magnetic moments from
M{\"o}ssbauer hyperfine fields~\cite{DUB09} it was deduced that the effective
hyperfine coupling constants for the two iron sites in Fe$_{3-y}$Co$_{y}$Si
alloys may differ by a factor of up to two in the composition range considered
here. As this deduction was based on the analysis of the neutron diffraction
data by Niculescu {\it et al.}~\cite{NIC79}, who assumed a unique cobalt moment, the
large difference in the $A$ values between the two sites is possibly an artifact.
Although the present estimates of the local moments are in reasonable agreement
with experimental and theoretical data, it cannot be excluded that due to
polarization contributions to $B_{\rm{hf}}$ from the environment of the Fe sites
the $A$ values are not completely independent from composition. This may modify
somewhat the detailed values of the moments and the assignment of the small
$B_{\rm{hf}}$ variation of component~I to a moment modulation at the Fe $4a$
sites may not be unambiguous.

\begin{table*}[htb]
\caption{\label{tab:mb}
         Results from the magnetic measurements and $^{57}$Fe M{\"o}ssbauer spectroscopy on Co$_{1+x}$Fe$_{2-x}$Si
         related to the composition $x$ and the number of valence electrons $N_v$.
         The values for the average hyperfine fields of the distributions $\left <B_{\rm{hf}} \right>$
         and for the hyperfine fields with the most probable $B_{\rm{hf}}$ are given in Tesla
         and have an error of $\pm 0.1$~T. The isomer shift $IS$ is given in mm/s with an uncertainty
         of $\pm 0.001$~mm/s. For labelling the two different iron sites we use the Wyckoff positions
         of the $x~=~1$ compound Co$_2$FeSi though due to replacement of Co by Fe the actual symmetry is lowered.
         Fe($8c$)$_{area}$ corresponds to the Fe fraction at the former $8c$ sites obtained from the M{\"o}ssbauer spectra.
         Fe($8c$)$_{calc.}$ is the corresponding Fe fraction calculated from stoichiometry.
         The magnetic moment per formula unit $m$ has a standard deviation of 1.5\%.
         The relative error of the determined Curie temperature is 0.4\%.}
\centering
\begin{ruledtabular}
\begin{tabular}{ccccccccccccccc}
$x$ & $N_v$ & \multicolumn{3}{c}{Fe($4a$)} & \multicolumn{3}{c}{Fe($8c$)} & Fe($8c$)$_{area}$ & Fe($8c$)$_{calc.}$ & $m_{\rm{Fe}(4a)}$ & $m_{\rm{Fe}(8c)}$ & $m$ & $m_{\rm{Co}}$ & $T_C$\\
 & &  $IS$ & $<$$B_{\rm{hf}}$$>$ & $B_{\rm{hf}}$ & $IS$ &$<$$B_{\rm{hf}}$$>$ & $B_{\rm{hf}}$ & (\%) & (\%) & ($\mu_{\rm B}$) & ($\mu_{\rm B}$) & ($\mu_{\rm B}$) & ($\mu_{\rm B}$) & (K)  \\
\hline
0 & 29.0 & 0.093 & 32.2 & 32.8 & 0.250 & 19.5 & 19.6 & 49.1 & 50.0 & 2.62 & 1.57 & 5.20 & 1.01 & 1036\\
0.2 & 29.2 & 0.099 & 32.9 & 33.3 & 0.250 & 19.4 & 19.6 & 44.3 & 44.4 & 2.66 & 1.57 & 5.60 & 1.40 & 1052\\
0.4 & 29.4 & 0.109 & 33.0 & 33.6 & 0.252 & 19.4 & 19.6 & 38.5 & 37.0 & 2.69 & 1.57 & 5.75 & 1.51 & 1059\\
0.5 & 29.5 & 0.113 & 33.4 & 33.7 & 0.252 & 19.6 & 19.7 & 34.6 & 33.3 & 2.70 & 1.58 & 5.62 & 1.42 & 1069\\
0.6 & 29.6 & 0.118 & 33.0 & 33.3 & 0.245 & 19.3 & 19.6 & 31.0 & 28.5 & 2.66 & 1.57 & 5.75 & 1.54 & 1058\\
1    & 30.0  & 0.125 & 31.1 & 31.3 & -      & -      & -      & 0     & 0      & 2.50 & -    & 5.92 & 1.71 & 1045\\
\end{tabular}
\end{ruledtabular}
\end{table*}

%%%%%%%%%%%%%%%%%%%%%%%%%%%%%%%%%%%%%%%%%%%%%%%%%%%%%%%%%%%%%%%%%%%%%%%%%%%%%%%
\section{Discussion}
\label{sec:discussion}

Within the Co$_{1+x}$Fe$_{2-x}$Si series of Heusler compounds
($0 \leq x \leq 1$) the valence electron concentration is changed from 29
for Co$_2$FeSi to 30 for CoFe$_2$Si. The Curie temperatures vary between 1036 and 1069~K.
The highest $T_C$ is obtained for the
composition Co$_{1.5}$Fe$_{1.5}$Si. 
The dependence on the composition can be explained by examination of the
atomic environments. It was found in earlier work~\cite{CoHMF2} that the $T_C$ of
Co-containing Heusler compounds with different magnetic sublattices depends
on the corresponding sublattice transition temperatures. These sublattice transition temperatures are
proportional to the spin moments $S$ and the Heisenberg exchange integral $J$
between the different sites. These variables are clearly modified with the
change in composition and thus the local environments.

As suggested by the analysis of the M{\"o}ssbauer data, the increase in the
total magnetic moment with increasing $N_v$ is mainly reflected in an increase
in the local moment of the cobalt atoms. $B_{\rm{hf}}$ and $m_{\rm{Fe}}$ remain
constant for Fe on former $8c$ sites since their nearest neighbor environment
(eight Fe atoms on the $4a$ positions) remains unchanged. On the contrary, it is
noteworthy that $T_C$ shows a comparable non-monotonic variation as the
hyperfine field for the Fe $4a$ sites (Figure~\ref{fig:Tc}). This $B_{\rm{hf}}$
variation is associated with the variation in the local environment of the Fe
atoms located on the $4a$ positions. Their nearest neighbor (NN) Co atoms are
consecutively substituted by Fe atoms which introduces disorder. The variation
in $B_{\rm{hf}}$ may reflect a variation in the local magnetic moments of iron
located on $4a$ sites. In that case, exchange interactions involving these sites
are modified as well.

For the Co$_{1+x}$Fe$_{2-x}$Si series, different exchange coupling integrals
need to be considered: $J_{\rm{Co-Co}}$, $J_{\rm{Fe-Fe}}$ and $J_{\rm{Co-Fe}}$.
In perfectly ordered regular Heusler type Co$_2$FeSi, $J_{\rm{Fe(4a)-Co(8c)}}$
(NN) and $J_{\rm{Co(8c)-Co(8c)}}$ (next nearest neighbors, NNN) occur. There are
no Fe--Fe interactions present. In the ordered inverse Heusler structure of
CoFe$_2$Si, $J_{\rm{Fe(4a)-Co(4c)}}$ and $J_{\rm{Fe(4a)-Fe(4d)}}$ are present in
the NN shells and $J_{\rm{Fe(4d)-Co(4c)}}$ occurs between NNN. No Co--Co
interactions arise in the first two atomic shells. In the off-stoichiometric
compounds, all combinations of Heisenberg exchange integrals are present.
Additionally, in the disordered structures with $x\neq 1$, Fe atoms located on
the former $8c$ positions may couple with Fe atoms on equal crystallographic
sites. Thus, only in the disordered samples additional $J_{\rm{Fe(8c)-Fe(8c)}}$
occurs between NNN. It seems likely that this additional Heisenberg exchange
integral plays an important role on the $T_C$ in the off-stoichiometric samples.

The effect of disorder on $T_C$ in this system is surprising. Generally,
disorder tends to reduce $T_C$, with the most extreme case being amorphous
transition metal alloys, which lack short and long range chemical and structural
order. These exhibit a reduced Curie temperature in comparison to their
crystalline analog~\cite{Coey}. Remarkably, in this system changes in the site
occupancy actually increase $T_C$. The modulation of the exchange energies due
to variations in magnetic moments and exchange integrals are the likely origin
for this behavior.

%%%%%%%%%%%%%%%%%%%%%%%%%%%%%%%%%%%%%%%%%%%%%%%%%%%%%%%%%%%%%%%%%%%%%%%%%%%%%%%
\section{Conclusion}

A series of Heusler compounds of the composition Co$_{1+x}$Fe$_{2-x}$Si 
($0 \leq x \leq 1$) with high Curie temperatures was synthesized to study 
trends in $m$ and $T_C$ in a small range of valence electron concentrations.
$^{57}$Fe M{\"o}ssbauer spectroscopy confirmed that in the off-stoichiometric
compounds the additional Fe atoms replace the Co atoms at the $8c$ positions
successively causing L2$_{1b}$-type disorder.

The total magnetic moment $m$ increases with increasing $N_v$ comparable to
variations in the local magnetic moment of cobalt derived from the M{\"o}ssbauer
data. A maximum Curie temperature of 1069~K is observed for the composition with $x=0.5$.
The hyperfine fields corresponding to iron on the $4a$ sites follow a trend
comparable to that of the $T_C$, which may reflect a similar variation in the
local magnetic moments of Fe atoms located on these positions. The $B_{\rm{hf}}$
of iron atoms located on the former $8c$ positions is constant with composition
reflecting a well-ordered nearest neighbor environment (Fe(4a)--Si(4b)).

In addition, the variations in the site occupancy and changes in the local
atomic environments modify the corresponding Heisenberg exchange integrals. Both
factors, local magnetic moments and exchange integrals, must be examined
carfeully to understand the trend in the $T_C$. Theoretical analysis is highly
desirable for an in-depth understanding of this behavior.

%%%%%%%%%%%%%%%%%%%%%%%%%%%%%%%%%%%%%%%%%%%%%%%%%%%%%%%%%%%%%%%%%%%%%%%%%%%%%%%
\bigskip
\begin{acknowledgments}

The authors gratefully acknowledge Lukas Wollmann for helpful discussions,
Walter Schnelle for performing magnetic measurements, Silvia Kostmann and Monika
Eckert for the microstructural analysis, Horst Borrmann for the X-Ray
diffraction measurements and Ulrike Schmidt and Anja V{\"o}lzke for chemical
analysis. This work was financially supported by the ERC Advanced Grant no. 291472
\textit{Idea Heusler}.

\end{acknowledgments}

%%%%%%%%%%%%%%%%%%%%%%%%%%%%%%%%%%%%%%%%%%%%%%%%%%%%%%%%%%%%%%%%%%%%%%%%%%%%%%%
% Create the reference section using BibTeX:
\bibliography{bibtexfile}

\end{document}